\documentclass[conference,final,10pt]{IEEEtran}
\usepackage[utf8]{inputenc}
\usepackage{caption,subcaption}
\usepackage{graphicx}
\usepackage{amsmath,amssymb,amsthm,mathtools}
\usepackage{color}
\usepackage{algorithm, algorithmic}
\usepackage{upgreek}
\usepackage{ragged2e}
\usepackage{psfrag}
\usepackage{stfloats}
\usepackage[noadjust]{cite}
\usepackage[draft,hidelinks]{hyperref}
\usepackage[nocomma]{optidef}
\usepackage{arydshln} % Needed for \hdashline

\usepackage[acronym,shortcuts]{glossaries}
\newacronym{GEV}{GEV}{Generalized Extreme Values}
\newacronym{EV}{EV}{Extreme Values}
\newacronym{CS}{CS}{Compressed Sensing}
\newacronym{CSI}{CSI}{channel state information}
\newacronym{BP}{BP}{basis pursuit}
\newacronym{OMP}{OMP}{orthogonal matching pursuit}
\newacronym{BPDN}{BPDN}{basis pursuit denoising}
\newacronym{RIP}{RIP}{restricted isometry property}
\newacronym{ETF}{ETF}{equiangular tight frame}
\newacronym{UNTF}{UNTF}{unit-norm tight frames}
\newacronym{SIDCO}{SIDCO}{sequential iterative decorrelation via convex optimization}
\newacronym{H-SIDCO}{H-SIDCO}{hungarian pairing SIDCO}
\newacronym{C-SIDCO}{C-SIDCO}{complex SIDCO}
\newacronym{QC-SIDCO}{QC-SIDCO}{quadratic complex SIDCO}
\newacronym{SVD}{SVD}{singular value decomposition}
\newacronym{mmWave}{mmWave}{millimeter wave}
\newacronym{MIMO}{MIMO}{multiple-input multiple-output}
\newacronym{AoA}{AoA}{angles of arrival}
\newacronym{AoD}{AoD}{angles of departure}
\newacronym{BS}{BS}{base station}
\newacronym{UE}{UE}{user equipment}
\newacronym{AWGN}{AWGN}{additive white Gaussian noise}
\newacronym{NMSE}{NMSE}{normalized mean square error}
\newacronym{PTF}{PTF}{Parseval tight frame}
\newacronym{PDF}{PDF}{probability density function}

\renewcommand{\smallskip}{\vspace{0.25cm}}
\newcommand{\bv}[1]{{\bf{#1}}}

\newcommand{\figsrootdir}{}
\newcommand{\figsdir}{./\figsrootdir/}

\newcommand{\referencesrootdir}{./}
\newcommand{\myreferences}{\referencesrootdir/listofpublications.bib}

\begin{document}

\title{A Frame-Theoretic Scheme for\\ Robust Millimeter Wave Channel Estimation}

\author{
\IEEEauthorblockN{Razvan-Andrei Stoica$^\dagger$, Giuseppe Thadeu Freitas de Abreu$^{\dagger,\ddagger}$ and Hiroki Iimori$^{\ddagger}$}
\IEEEauthorblockA{
$^\dagger$ Focus Area Mobility, Jacobs University Bremen, Campus Ring 1, 28759 Bremen, Germany \\
Email: {\tt [r.stoica,g.abreu]@jacobs-university.de}\\
$^\ddagger$ Department of Electrical and Electronic Engineering, Ritsumeikan University, Kusatsu, Shiga, 525-8577 Japan\\
Email: {\tt [g-abreu@fc,h.iimori@gabreu.se].ritsumei.ac.jp}}\\[-5ex]
}

% \thanks{Manuscript received: ...~ . This work has been supported by the EU project HIGHTS, number 636537, program H2020. } 
%
% \thanks{This work has been supported by the EU project HIGHTS, number 636537, program H2020.}
%
% \thanks{Razvan-Andrei Stoica and Dr. Stefano Severi are with the Focus Area Mobility, Jacobs University Bremen, Bremen, Germany.
% (e-mail: {\tt [r.stoica,s.severi]@jacobs-university.de}).}
%
% \thanks{Prof. Dr. Giuseppe Thadeu Freitas de Abreu is with both the Focus Area Mobility, Jacobs University Bremen, Bremen, Germany, and with the Department of Electrical and Electronic Engineering, Ritsumeikan University, Kusatsu, Shiga, Japan. (e-mail: {\tt g.abreu@jacobs-university.de}, {\tt g-abreu@fc.ritsumei.ac.jp}).}}

\maketitle

\begin{abstract} 
We propose a new scheme for the robust estimation of the \ac{mmWave} channel.
Our approach is based on a sparse formulation of the channel estimation problem coupled with a frame theoretic representation of the sensing dictionary.
To clarify, under this approach, the combined effect of transmit precoders and receive beamformers is modeled by a single frame, whose design is optimized to improve the accuracy of the sparse reconstruction problem to which the channel estimation problem is ultimately reduced.
The optimized sensing dictionary frame is then decomposed via a Kronecker decomposition back into the precoding and beamforming vectors used by the transmitter and receiver.
Simulation results illustrate the significant gain in estimation accuracy obtained over state of the art alternatives.
As a bonus, the work offers new insights onto the sparse \ac{mmWave}-\ac{MIMO} channel estimation problem by casting the trade-off between correlation and variation range in terms of frame coherence and tightness.

\end{abstract}

\glsresetall

%\vspace{-1ex}
\section{Introduction}
%\vspace{-1ex}
%
The increasing demands in terms of higher rate, more access and lower latency at the physical link, coupled with the lack of spectral resources in conventional cellular systems is recently strongly motivating the development of \ac{mmWave} communications \cite{RappaportTamir2013}.
%
% Insofar \ac{mmWave} it is currently seen as one of the key enabler of multiple gigabit wireless access in the upcoming iterations of cellular networks \cite{RappaportTamir2013}.

In principle, the larger bandwidths and shorter wave lengths of \ac{mmWave} systems enable the provision of higher communication rates \cite{Zorzi2015}, and respectively, the equipping of larger antennas arrays at transmitters and receivers favors the utilization of \ac{MIMO} architectures \cite{GaoHeath2016, SunRappaport2014}.

In practice, however, hardware costs and other implementation issues challenge the realization of \ac{mmWave} systems, which therefore must be counter-acted via dedicatedly designed signal processing methods \cite{ZhouOhashi2014, LiangDong2014}.

In turn, previous work has demonstrated \cite{Heath_ChEst2014, Lee2014Globecomm, GhauchEst2015, RialITA2015, Montagner2015, NiDong2016} that the efficacy of signal processing in ameliorating the hardware challenges of \ac{mmWave} systems depends highly on the quality of \textit{\ac{CSI}}.
In fact, although hybrid precoding with \textit{partial \ac{CSI}} has been well studied \cite{MyListOfPapers:HeathJSASP2016}, the substantial performance losses resulting from imperfect/partial \ac{CSI} only further motivate the quest for better methods for channel acquisition \cite{NiDong2016, Montagner2015, GhauchEst2015, Heath_ChEst2014, RialITA2015, Lee2014Globecomm}.

Retrieving complete and accurate \ac{mmWave} \ac{CSI} is challenging in practice due to the rapid variation and severe path-loss experienced under the high operating frequencies.

In answer to this challenge, a sparse formulation of the \ac{mmWave} \ac{MIMO} channel estimation problem was proposed in \cite{Heath_ChEst2014} which allowed the use of \ac{CS} \cite{DonohoCS2006} for the scant channel recovery problem \cite{RialITA2015}, which posteriorly was improved by the introduction of a greedy \ac{OMP} recovery algorithm \cite{TroppOMP2007}.

Recognizing that the efficacy of \ac{OMP} in noisy systems is limited, as the method fails to exactly fit linear systems, an alternative solver to mitigate this problem has been proposed in our previous work \cite{malla2016iswcschannel} in which the \ac{BPDN} \cite{donoho1994basispursuit} has been leveraged as a more efficient solution.
Furthermore, in \cite{malla2016iswcschannel} the sparsity of the problem was enhanced through a reweighted $\ell_1$-minimization formulation \cite{CandesBoydL12007}, which led to an efficient iterative \ac{BPDN}$-\ell_1$ sparse recovery.

In this paper, we continue this trend and further contribute with a technique for joint channel estimation and training beamformer optimization.
The generic optimization of training vectors is performed based on Frame Theory and its applicability to sparse recovery, \cite[Ch.9]{casazza2012finite}, \cite{rusu2016designing}. 
In turn, the measurement matrix selection problem is solved by a decoupled, flexible low-coherence tight frame design, with increased robustness compared to conventional random or optimized training vectors \cite{chen2013projectiondesignframescs}.
%
%The solution resumes thus to designing a complex low-coherence tight frame as the natural extension of the real domain \ac{SIDCO} scheme proposed in \cite{rusu2016designing}.

In the remainder of the paper the following notation is used:
\begin{itemize}
\item $\bv{X}$, $\bv{x}$ and $x$ represent a matrix, a vector, and a scalar;
\item $\|\bv{X}\|_\text{F}, \|\bv{x}\|_2$ and $\|\bv{x}\|_{\infty}$ denote the Frobenius, Euclidean and $\infty$-norms;

\item $\bv{X}^{\text{T}}, \bv{X}^{\text{H}}\,\, \text{and}\,\, \bv{X}^{\text{*}}$ denote the transpose, complex conjugate transpose and conjugate of matrix $\bv{X}$;
\item $\bv{X}\otimes\bv{Y}$ is the Kronecker product of $\bv{X}$ and $\bv{Y}$;
\item $\rm{diag}(\bv{x})$ denotes the diagonal matrix with diagonal $\bv{x}$;
\item $\rm{vec}(\bv{X})$ is a column vector with all columns of $\bv{X}$ stacked;
%
% and the inverse operation is noted as $\rm{ivec}(\bv{X})$.
%
\item $\bv{I}_N$ and $\bv{0}_N$  denote the $N$-sized identity and null matrices.
\end{itemize}

\vspace{-1ex}
\section{Problem Formulation}
\subsection{\hspace{-1ex} Millimeter\! Wave\! Channel\! Estimation\! \&\! Compressive\! Sensing}

A downlink \ac{MIMO} \ac{mmWave} system formed of a \ac{BS} with $T$ transmit antennas and an \ac{UE} with $R$ receive antennas is considered.
It is also assumed that the \ac{BS} uses $M_\text{T}$ training beamforming vectors to transmit a known training signal $\bv{S}$, while the \ac{UE} applies $M_\text{R}$ combining vectors for each beamforming one in order to estimate the channel $\bv{H} \in \mathbb{C}^{R\times T}$.

It follows that the receive signal matrix at the \ac{UE}, denoted by $\bv{Y}\in\mathbb{C}^{M_\text{R}\times M_\text{T}}$, is given by
\begin{equation}
\label{eq_ReceivedMatrix}
\bv{Y} = \bv{V}^{\text{H}} \bv{H} \bv{U} \bv{S} + \bv{N},
\end{equation}
where the precoding and combining (TX/RX beamforming) matrices are given by $\bv{U}\triangleq[\bv{u}_1,\cdots,\bv{u}_{M_\text{T}}] \in \mathbb{C}^{T\times M_\text{T}}$ and  $\bv{V}\triangleq[\bv{v}_1,\cdots,\bv{v}_{M_\text{R}}] \in \mathbb{C}^{R \times M_\text{R}}$, respectively, $\bv{N} \in \mathbb{C}^{M_\text{R}\times M_\text{T}}$ denotes circularly symmetric complex \ac{AWGN}, and $\bv{H}$ is the \ac{mmWave} channel matrix.

Following the usual sparse multipath scatter channel model \cite{GaoHeath2016,NiDong2016,Heath_ChEst2014,GhauchEst2015,Montagner2015,Lee2014Globecomm,RialITA2015}, we may rewrite $\bv{H}$ as
\begin{equation}
\label{channel_model}
\bv{H} = \sqrt{\frac{TR}{L}}\sum_{l=1}^{L}\gamma_l \bv{a}_\text{r}(\phi_l^\text{r}) \mathbf{a}_\text{t}^{\textup{H}}(\phi_l^\text{t}),
\vspace{-1ex}
\end{equation}
where $L$ is the number of propagation paths, $\gamma_l\sim\mathcal{CN}(0,\sigma_{\gamma}^2)$ is the complex gain of the $l$-th path, and $\bv{a}_\text{r}(\phi_l^\text{r})$ and $\mathbf{a}_\text{t}(\phi_l^\text{t})$ are the array response vectors respectively at the receiver and transmitter, with corresponding \ac{AoA} and \ac{AoD} denoted by $\phi_l^\text{r},\; \phi_l^\text{t}\in[0,2\pi]$.

The channel matrix described by equation \eqref{channel_model} can also be more compactly expressed as
\begin{equation}
\label{eq_channelMatrix}
\bv{H} = \bv{A}_\text{R} \bv{H}_{\gamma} \bv{A}_\text{T}^{\!\textup{H}},
\end{equation}
with $\bv{A}_\text{R} \!\triangleq\! [\bv{a}_\text{r}(\phi_1^\text{r}),\cdots,\bv{a}_\text{r}(\phi_L^\text{r})]$, $\bv{A}_\text{T} \!\triangleq\! [\bv{a}_\text{t}(\phi_1^\text{t}),\cdots,\bv{a}_\text{t}(\phi_L^\text{t})]$, and $\bv{H}_{\gamma} \!\triangleq\! \sqrt{\frac{TR}{L}}\,\rm{diag}(\gamma_1,\cdots,\gamma_L)$.

For the sake of simplicity, identity signaling is assumed hereafter, such that the training transmit symbol matrix is given by $\bv{S} = \bv{I}_{M_\text{T}}$, which in turn implies that equation \eqref{eq_ReceivedMatrix} can be rewritten in a vectorized form as
\begin{equation}
\label{eq_ReceiveVector}
\bv{y} \triangleq \rm{vec}(\bv{Y}) \!=\! (\bv{U}^{\text{T}} \!\otimes\! \bv{V}^{\text{H}}) (\bv{A}_\text{T}^{*}\!\otimes\! \bv{A}_\text{R}){\rm{vec}}(\bv{H}_{\gamma}) \!+\! \rm{vec}(\bv{N}),
\end{equation}
where $\bv{y}\in\mathbb{C}^{M_\text{R}M_\text{T}\times 1}$.

A sparse characterization of equation \eqref{eq_ReceiveVector} can be obtained  as follows \cite{Heath_ChEst2014,Lee2014Globecomm}.
First, consider expanded versions of the scatter matrices $\bv{A}_\text{R}$ and $\bv{A}_\text{T}$ defined by $\hat{\bv{A}}_\text{R} \!\!\triangleq\!\! [\bv{a}_\text{r}(\theta_0),\cdots,\bv{a}_\text{r}(\theta_{G_\text{R}-1})]$ and $\hat{\bv{A}}_\text{T} \triangleq [\bv{a}_\text{t}(\theta_0),\cdots,\bv{a}_\text{t}(\theta_{G_\text{T}-1})]$ where the \acp{AoD} and \acp{AoA} $\theta_{{g_\text{t}}}$ and $\theta_{{g_\text{r}}}$ lay on a sufficiently fine quantization grid, $i.e.$
\begin{equation}
\label{eq_Discretization}
\theta_{\!{g_\text{t}}} \triangleq \frac{2\pi g_\text{t}}{G_\text{T}},\quad \text{and} \quad \theta_{\!{g_\text{r}}} \triangleq \frac{2\pi g_\text{r}}{G_\text{R}},
\end{equation}
with $g_\text{t} \!=\! \{0,\dots,G_\text{T}\!-\!1\}$, $g_\text{r} \!=\! \{0,\dots,G_\text{R}\!-\!1\}$, and respectively,
$(G_\text{T}, G_\text{R}) \!\gg\ \!\! L$.

Next, expand also $\bv{H}_\gamma$ into a sparse matrix $\hat{\bv{H}}_\gamma$, whose only non-zero entries are the $L$ elements satisfying
\begin{equation}
[\hat{\bv{H}}_\gamma]_{i,j} = \gamma_\ell \quad \Longleftrightarrow \quad
\left\{
\begin{array}{l}
\|\theta_{\!i} - \phi_\ell^\text{r}\|_2 < \|\theta_{\!g_\text{r}\neq i} - \phi_\ell^\text{r}\|_2, \\[1ex]
\|\theta_{\!j} - \phi_\ell^\text{t}\|_2 < \|\theta_{\!g_\text{t}\neq i} - \phi_\ell^\text{t}\|_2,
\end{array}
\right.
\end{equation}
for every $\ell=\{1,\cdots,L\}$.

%Then, we may construct
%%
%\begin{equation}
%\label{eq_ChannelSparse}
%\hat{\bv{H}} \approx \hat{\bv{A}}_\text{R} \hat{\bv{H}}_\gamma\hat{\bv{A}}^{\!\text{H}}_\text{T}.
%\end{equation}

And finally, obtain \cite{malla2016iswcschannel}
\begin{equation}
\label{eq_ChannelSparse2}
\bv{y}  = \underbrace{(\bv{U}^{\text{T}} \otimes \bv{V}^{\text{H}})}_{\bv{\Phi}}
\underbrace{(\hat{\bv{A}}_\text{T}^{*}\otimes \hat{\bv{A}}_\text{R})}_{\bv{\Psi}}
\underbrace{{\rm{vec}}(\hat{\bv{H}}_\gamma)}_{\bv{x}} + \underbrace{\rm{vec}(\bv{N})}_{\bv{n}},
\end{equation}
where the \emph{measurement matrix} $\bv{\Phi}\!\triangleq\! {(\bv{U}^{\text{T}}\! \otimes\! \bv{V}^{\text{H}})} \in \mathbb{C}^{M_\text{T}M_\text{R} \times TR}$, the \emph{sparse dictionary} $\bv{\Psi}\triangleq{(\hat{\bv{A}}_\text{T}^{*}\otimes \hat{\bv{A}}_\text{R})} \in \mathbb{C}^{TR\times G_\text{T}G_\text{R}}$, the $L$-\emph{sparse vector} $\bv{x}\triangleq{\rm{vec}}(\hat{\bv{H}}_\gamma) \in \mathbb{C}^{G_\text{T}G_\text{R}\times 1}$ and the noise vector $\bv{n}$ have been implicitly defined.

% \newpage

\subsection{Previous Contributions}

Under the assumption that the \ac{AoA} and \ac{AoD}
angles $\phi_l^\text{r}$ and $\phi_l^\text{t}$ are known\footnote{As literature on \ac{AoA} estimation is vast we refrain from further discussion.}, and in light of the model expressed by equation \eqref{channel_model}, the mmWave channel estimation problem amounts to estimating the complex gains $\{\gamma_1,\cdots,\gamma_L\}$. %, \gamma_l\sim\mathcal{CN}(0,\sigma_{\gamma}^2)$.
And under the further assumption that the precoding and combining matrices $\bv{U}$ and $\bv{V}$ are given, the vectorized expression of equation \eqref{eq_ChannelSparse2} enables the mmWave channel estimation problem to be solved as the sparse recovery \ac{CS} optimization problem:
\begin{subequations}
\begin{eqnarray}
\hspace{-1ex}
&\underset{\bv{x}} {\mathrm{minimize}}& \|\bv{x}\|_{{\ell}_0}, \\
& \mathrm{subject \,\,to} & \bv{y} = \underbrace{\bv{\Phi \Psi}}_{\bv{\Omega}}\bv{x} + \bv{n},
\end{eqnarray}
\label{eq_l0combinatorial}%
\end{subequations}
where we have explicitly identified the \emph{equivalent sensing matrix} $\bv{\Omega}\in\mathbb{C}^{M_\text{T}M_\text{T}\times G_\text{T}G_\text{R}}$ for future convenience.

The problem formulated above can be solved via the \ac{OMP} \cite{TroppOMP2007} algorithm, \textit{e.g.} as proposed in \cite{Heath_ChEst2014} and \cite{RialITA2015}.
More recently, we have shown in \cite{malla2016iswcschannel} that the latter can be enhanced by relaxing the problem \eqref{eq_l0combinatorial} to the associated $\ell_1$-norm formulation
\vspace{-2ex}
\begin{subequations}
\begin{eqnarray}	
&\underset{\bv{x}} {\mathrm{minimize}} & \|\bv{x}\|_{{\ell}_1}, \\
& \mathrm{subject \,\,to} & \| \bv{y} - \bv{\Omega x}\|_{\ell_2} \leq \delta,
\end{eqnarray}
\label{eq_bpdn}%
\end{subequations}
which can then be solved via \ac{BPDN}, thus mitigating the noisy recovery limitations encountered by classical \ac{OMP}.

In fact, the problem, can be even more accurately solved if the \ac{BPDN} solver is further combined with the sparsity-enhancing iterative $\ell_1$-reweighing scheme of \cite{CandesBoydL12007}, as also shown in \cite[Alg. 1]{malla2016iswcschannel}.

Another approach to further improve the performance of mmWave channel estimation that received comparatively less attention so far is to optimize the sensing matrix $\bv{\Omega}$, given a certain discrete angle dictionary $\bv{\Psi}$, which is usually fixed by means of hardware/processing requirements.

Deriving a method to optimize $\bv{\Omega}$ given $\bv{\Psi}$, which in turn reduces to optimizing $\bv{\Phi}$, is the objective and the main contribution of this article, and the focus of the next section. 

\section{Frame-theoretical Design of\\ Precoding \& Beamforming Matrices}
\label{sect:frames_basics}
\ac{CS} is a direct application of a larger framework of linear projections, namely Frame Theory \cite[Ch. 9]{casazza2012finite}, \cite[Sect. 7.2]{kovacevic2008introduction}.
In a general sense, a \textit{frame} is defined as a set of $N$ vectors $\bv{F} \triangleq \left[\bv{f}_1, \bv{f}_2, \dots, \bv{f}_N \right]$ over a Hilbert space $\mathbb{H}^M$ (reduced to $\mathbb{C}^M$ in the current setup) with $M < N$ and
%
% \vspace{-1ex}
\begin{eqnarray}
\label{eq:frame_defi}
\alpha \|\bv{v}\|_2^2 \leq \|\bv{F}^{\text{H}} \bv{v}\|_2^2 \leq \beta \|\bv{v}\|_2^2,
\vspace{-1ex}
\end{eqnarray}
where ($\alpha,\beta$), $0 < \alpha \leq \beta < \infty$, are the finite highest lower and lowest higher frame bounds, respectively \cite{casazza2012finite}.

A frame is \textit{tight} iff $\alpha = \beta$, and \textit{unit-norm} iff $\|\bv{f}_i\|_2 = 1,\; \forall i$.
\textit{\Ac{UNTF}} have both these properties so that $\alpha = \beta = \frac{N}{M} \triangleq \rho(\bv{F})$, where $\rho(\bv{F})$ is known as the \emph{redundancy} of the frame \cite[Ch. 1]{casazza2012finite}.

Similar to the \ac{RIP} in \ac{CS}, the bounding property expressed by equation \eqref{eq:frame_defi} offers a measure of how close a frame is to an orthogonal basis with respect to any projected vector in the spanned space.
But another measure of the frame's similarity to an orthonormal basis is its \textit{mutual coherence}, defined (for a unit-norm frame) as
\vspace{-0.75ex}
\begin{equation}
\label{eq:mutual_coh_defi}
 \mu(\bv{F}) \triangleq \max_{i \neq j} \| \bv{f}_i^{\text{H}}\bv{f}_j \|_2 = \!\!\! \max_{\bv{G} \triangleq \bv{F}^H\bv{F}, i\neq j} |g_{ij}| \geq \sqrt{\tfrac{N-M}{M(N-1)}},
\vspace{-0.5ex}
\end{equation}
where $\bv{G}$ is known as the \emph{Gram operator}, and the lower bound on the right-hand side is the \textit{Welch bound} \cite{strohmer2003grassmanianapps} for $N \leq M^2$.

The performance of pursuit algorithms can be studied under concepts like coherence \cite{donoho2006compressed} and the restricted isometry property \ac{RIP} \cite{tropp2006justrelax}.
As a result, there are two general requirements on the design of the measurement matrix $\bv{\Phi}$:
\begin{enumerate}
 \item $\bv{\Phi}$ must be highly incoherent in order to preserve the salient information of sparse vectors;
 \item $\bv{\Omega}$ must satisfy the \ac{RIP} in order to afford  robustness to the reconstruction.
\end{enumerate}

We may remark at this stage that the dictionary $\bv{\Psi}$ in equation \eqref{eq_ChannelSparse2} can in fact be identified as a \emph{harmonic frame}, sampled out of the discrete Fourier matrix of size $G_\text{T} \times G_\text{R}$, so that $\bv{\Psi}$ is an \ac{UNTF} by construction \cite{kovacevic2008introduction}.
Also interesting to notice is the fact that this frame admits a natural Kronecker-decomposable form $\bv{\Psi}\triangleq{(\hat{\bv{A}}_\text{T}^{*}\otimes \hat{\bv{A}}_\text{R})}$, as seen before.

In light of all the above, our goal is to design the optimized measurement matrix $\bv{\Phi}$ as a \emph{Kronecker-decomposable}, \textit{normalized tight frame} with \textit{low-coherence} and with a \emph{\ac{RIP}-compliant} associated sensing matrix $\bv{\Omega}=\bv{\Phi}\bv{\Psi}$, which is addressed in the sequel.

\subsection{\!\!\!QC-SIDCO:\! Measurement Matrix as a Low-coherence\! Frame}

A low-coherence frame can be generated from a given unit-norm frame $\tilde{\bv{F}} \in \mathbb{C}^{M \times N}$ by iteratively decorrelating its vectors while constraining the feasibility region to an $M$-ball.
This scheme, referred to as \ac{SIDCO}, was originally introduced in \cite{rusu2016designing} only for frames in $\mathbb{R}^M$.
More recently, a strategy to generalize \ac{SIDCO} to frames in $\mathbb{C}^M$, referred to as \ac{C-SIDCO}, was reported by \cite{rusu2018algorithms}.
An explicit and complete mathematical formulation of \ac{C-SIDCO} was, however, not given in \cite{rusu2018algorithms}.
A variation of the latter based on an explicit quadratic program is offered below.

Consider an existent unit-norm frame $\tilde{\bv{F}} \!\!\in\!\! \mathbb{C}^{M \times N}$. 
The strategy of \ac{C-SIDCO} \cite{rusu2018algorithms} is to minimize mutual coherence by iteratively solving the problem 
\begin{subequations}
\begin{eqnarray}	
&\underset{\bv{f}_i \in \mathbb{C}^{M}} {\mathrm{minimize}} & \|\tilde{\bv{F}}_i^{\text{H}} \bv{f}_i\|_{\infty},\\
& \mathrm{subject \,\,to} & \|\bv{f}_i - \tilde{\bv{f}}_i\|_2^2 \leq T_i,
\end{eqnarray}
\label{csidco}%
\end{subequations}
for all $i$ vectors, where $\tilde{\bv{F}}_i$ denotes the $\tilde{\bv{f}}_i$-pruned existent frame and the search $M$-ball radius of vector $\bv{f}_i$ is given by
\begin{eqnarray}\label{eq:mballradius}
 T_i \leq 1 - \max_{j; j \neq i} |g_{ij}|^2,
\end{eqnarray}
so that $T_i$ is constrained to the largest $M$-ball such that the prospective solution $\bv{f}_i$ cannot be collinear with other $\tilde{\bv{f}}_j, j\neq i$.

In order to circumvent the additional challenge of optimizing in the complex domain, the space $\mathbb{C}^M$ is reinterpreted as $\mathbb{R}^{2M}$, with interlaced real and imaginary components.

The generically formulated \ac{C-SIDCO} approach described by equation \eqref{csidco} can be explicitly reformulated as the quadratic program
\begin{subequations}
\begin{eqnarray}	
&\hspace{-4ex} \underset{\bv{x} \triangleq \left[\bv{f}_i; t_\mathcal{R}; t_\mathcal{I}\right] \in \mathbb{R}^{2M+2}} {\mathrm{minimize}} & \bv{x}^{\text{T}}\bv{Q}\bv{x},\\
 &\hspace{-4ex} \mathrm{subject \,\,to} & \bv{A}_{\mathcal{R},1} \bv{x} \leq 0,\; \bv{A}_{\mathcal{R},2} \bv{x} \leq 0,\label{eq:stdf_real}\\
&			  & \bv{A}_{\mathcal{I},1} \bv{x} \leq 0,\; \bv{A}_{\mathcal{I},2} \bv{x} \leq 0,\label{eq:stdf_imag}\\
&			  & \bv{x}^{\text{T}}\bv{B}\bv{x} - 2\bv{b}^{\text{T}}\bv{x} + 1 - T_i \leq 0,\label{eq:mball_ctr}
\vspace{-1ex}
\end{eqnarray}
\label{csidco_solution}%
where
\begin{eqnarray}
&&\hspace{-4ex}\bv{Q} \triangleq 
\left[\begin{array}{c c} 
\bv{0}_{2M} & \bv{0}_{2M\times 2}\\
\bv{0}_{2\times 2M} & \bv{I}_{2}
\end{array}\right] \in \mathbb{R}^{(2M+2) \times (2M+2)},\\
&&\hspace{-4ex}\bv{A}_{\mathcal{R},1} \triangleq \left[\tilde{\bv{F}}_i^{\text{T}}\,\,\, -\!\bv{1}_{(\!N\!-\!1\!)\times1}\,\,\, \bv{0}_{(n\!-\!1)\times1}\right]\in \mathbb{R}^{(N-1)\times(2M\!+\!2)},\nonumber\\\
&&\hspace{-4ex}\quad\\
&&\hspace{-4ex}\bv{A}_{\mathcal{R},2} \triangleq \left[-\tilde{\bv{F}}_i^{\text{T}}\,\,\, -\bv{1}_{(N-1)\times1}\,\,\, \bv{0}_{(n-1)\times1}\right]\!\in\! \mathbb{R}^{(\!N\!-\!1\!)\times(2M\!+2)},\nonumber\\\
&&\hspace{-4ex}\quad\\
&&\hspace{-4ex}\bv{A}_{\mathcal{I},1} \triangleq \left[\tilde{\bv{F}}_i^{\text{T}}\bv{D}_M\,\,\, \bv{0}_{(N-1)\times1}\,\,\, -\bv{1}_{(n-1)\times1}\right]\!\in\! \mathbb{R}^{(\!N\!-\!1\!)\times(2M\!+2)},\nonumber\\\
&&\hspace{-4ex}\quad\\
&&\hspace{-4ex}\bv{A}_{\mathcal{I},2} \triangleq \left[-\tilde{\bv{F}}_i^{\text{T}}\bv{D}_M\,\,\, \bv{0}_{(N\!-1)\!\times\!1}\,\,\, -\!\bv{1}_{(n-1)\times1}\right]\!\in\! \mathbb{R}^{(\!N\!-\!1\!)\!\times\!(2M\!+2)}\!\!,\nonumber\\\
&&\hspace{-4ex}\quad\\
&&\hspace{-4ex}\bv{D}_M \triangleq 
\left[\begin{array}{c c} 
0 & -1\\
1 & 0
\end{array}\right] \otimes \bv{I}_M \in \mathbb{R}^{(2M+2)\times(2M+2)}\\
&&\hspace{-4ex}\bv{B} \triangleq 
\left[\begin{array}{c c} 
\bv{I}_{2M} & \bv{0}_{2M\times 2}\\
\bv{0}_{2\times 2M} & \bv{0}_{2\times 2}
\end{array}\right] \in \mathbb{R}^{(2M+2)\times(2M+2)},\\
&&\hspace{-4ex}\bv{b}^\text{T} \triangleq \left[\tilde{\bv{f}}_i^{\text{T}}\,\,\,0\,\,\,0\right] \in \mathbb{R}^{2M+2}.
\end{eqnarray}
\end{subequations}

We remark that thanks to the replacement of the linear program adopted in \cite[Alg. 1]{rusu2016designing} with the quadratic program given by equation \eqref{csidco_solution}, and the direct interlacing from $\mathbb{C}^M$ to $\mathbb{R}^{2M}$ achieved in our formulation via the matrices $\bv{D}_M,\bv{A}_{\mathcal{R},1},\bv{A}_{\mathcal{R},2},\bv{A}_{\mathcal{I},1}$ and $\bv{A}_{\mathcal{A},2}$, \ac{C-SIDCO} becomes here a simple extension of the original \ac{SIDCO} algorithm, which converges (absolutely) to the local minimum coherence points since the original concept of $M$-ball packings is preserved.

For all the above, the explicit quadratic reformulation of \ac{C-SIDCO} offered above is \textit{original}, and can be directly coded on top of optimized standard quadratic solvers, unlike the formulation presented in \cite{rusu2018algorithms}. 

We therefore refer to this realization of \ac{SIDCO} as the \ac{QC-SIDCO} algorithm.

\subsection{Beamformers: Decomposition of QC-SIDCO $\bv{\Phi}$}

The frame obtained by the method described above is not strictly tight, unless an \ac{ETF} is reached\footnote{\acp{ETF} are the closest frame analogies \!to orthogonal bases, attaining tightness and uniform lowest coherence, and therefore also the Welch bound \cite{strohmer2003grassmanianapps}.}, which rarely happens in practice, since \acp{ETF} exist only for particular dimensions.

Fortunately, tightness of the frame $\bv{F}$ constructed via the \ac{QC-SIDCO} frame method can be posteriorly enforced by applying polar decomposition \cite[Th. 2]{tropp2005designing}, which yields the \ac{UNTF} closest, in Frobenius-norm sense, \ac{QC-SIDCO} frame.
For details we refer the reader to \cite{tropp2005designing}.

Finally, in order to uniformly distribute the sensing cost of $\bv{\Omega}$, the polar-decomposed \ac{QC-SIDCO}-designed measurement matrix $\bv{\Phi}$ is obtained by normalizing the latter frame $\bv{F}^*$, $i.e.$
\begin{equation}\label{eq_optimizedPhi}
\bv{\Phi} = \frac{\sqrt{TR}\,\bv{F}^*}{\|\bv{F}^*\|_\text{F}}.
\end{equation}

Returning to our original problem of mmWave channel estimation, however, we remark that the measurement matrix obtained as explained above still needs to be decomposed into precoding and combining beamforming matrices $\bv{U}$ and $\bv{V}$ in order for the channel estimation method to be practically implementable.
This can be achieved by solving the problem
\begin{eqnarray}
\label{reconstruction_beamformers}
&\hspace{-4ex} \underset{\bv{U} \in \mathbb{C}^{T \times M_\text{T}}, \bv{V} \in \mathbb{C}^{R \times M_\text{R}}}{\mathrm{minimize}} & \lVert \bv{\Phi} - (\bv{U}^{\text{T}} \otimes \bv{V}^{\text{H}}) \rVert_\text{F},
\vspace{-1ex}
\end{eqnarray}
which reformulated by vectorization becomes
%
%
% IS THE LAST TRANSPOSISION CORRECT (in red) ??
%
%
\begin{eqnarray}
\label{reconstruction_beamformers_rank1}
&\hspace{-4ex} \underset{\bv{U} \in \mathbb{C}^{T \times M_\text{T}}, \bv{V} \in \mathbb{C}^{R \times M_\text{R}}}{\mathrm{minimize}} & \lVert {\boldsymbol{\varPhi}} - \left(\rm{vec}({\bv{U}^{\text{T}}}) \rm{vec}(\bv{V}^{\text{H}})^{\text{T}}\right) \rVert_\text{F},
\vspace{-1ex}
\end{eqnarray}
where 
\begin{eqnarray}
\label{measurement_perm}
\boldsymbol{\varPhi}^\text{T} \triangleq &&\\
&&\hspace{-9ex}
\left[
\begin{array}{@{}c@{\,}c@{\,}c@{\,}c@{\,}c@{\,}c@{\,}c@{}}
{\rm{vec}}({\bv{\Phi}}_{11}),&
\cdots,&
{\rm{vec}}({\bv{\Phi}}_{M_\text{T}1}),&
\cdots,&
{\rm{vec}}({\bv{\Phi}}_{1T}),&
\cdots,&
{\rm{vec}}({\bv{\Phi}}_{M_\text{T}T})
\end{array}
\right]\!\!,\nonumber
\end{eqnarray}
with
\begin{equation}
\label{orig_matrix_phi}
\bv{\Phi} = \left[
\begin{array}{c : c : c : c}
\bv{\Phi}_{11} & \bv{\Phi}_{12} & \cdots & \bv{\Phi}_{1T}\\
\hdashline
\bv{\Phi}_{21} & \bv{\Phi}_{22} & \cdots & \bv{\Phi}_{2T}\\
\hdashline
\vdots & \vdots & \vdots & \vdots\\
\hdashline
\bv{\Phi}_{M_\text{T}1} & \bv{\Phi}_{M_\text{T}2} & \cdots & \bv{\Phi}_{M_\text{T}T}\\
\end{array}
\right].
\end{equation}

Notice that the matrix
$\boldsymbol{\varPhi}$ defined above is a rectangular matrix, such that as a result of this reshaping, the solution of equation \eqref{reconstruction_beamformers_rank1} can be easily obtained via \ac{SVD}, yielding
\begin{equation}
\label{eq:FrameUV}
\rm{vec}(\bv{U}^T)= \sqrt{\sigma_{\boldsymbol{\varPhi}}} {\bv{u}}_{\boldsymbol{\varPhi}}\quad \text{and}\quad
\rm{vec}(\bv{V}^H) = \sqrt{\sigma_{{\boldsymbol{\varPhi}}}} {\bv{v}}_{\boldsymbol{\varPhi}},
\end{equation}
where ${\bv{u}}_{\boldsymbol{\varPhi}}$ and ${\bv{v}}_{\boldsymbol{\varPhi}}$ are the dominant left and right singular vectors, and $\sigma_{{\boldsymbol{\varPhi}}}$ the dominant singular value of ${\boldsymbol{\varPhi}}$.

With knowledge of the vectorized forms $\rm{vec}(\bv{U}^T)$, $\rm{vec}(\bv{V}^T)$ as in equation \eqref{eq:FrameUV}, the precoding and combining matrices $\bv{U}$ and $\bv{V}$  are finally obtained such that $\bv{\Phi}\!\triangleq\! {(\bv{U}^{\text{T}}\! \otimes\! \bv{V}^{\text{H}})}$, as desired for a practical implementation.

Given all the above, the proposed measurement matrix construction for application in the mmWave channel estimation problems described by equations \eqref{eq_l0combinatorial} and \eqref{eq_bpdn} can therefore be summarized as follows:
\begin{enumerate}
\item Generate a unit-norm low-coherence frame $\bv{F}$ through the \ac{QC-SIDCO} scheme described by equation \eqref{csidco_solution};
\item Apply \ac{SVD} and polar decomposition to $\bv{F}$, \cite[Th. 2]{tropp2005designing}, thus obtaining an \ac{UNTF} with low-coherence $\bv{F}$;
\item Obtain the ideal measurement matrix $\bv{\Phi}$ from eq. \eqref{eq_optimizedPhi};
\item Decompose $\bv{\Phi}$ using equation \eqref{eq:FrameUV}, and reshape $\rm{vec}(\bv{U}^T)\rightarrow \bv{U}$ and  $\rm{vec}(\bv{V}^H)\rightarrow \bv{V}$ accordingly for suboptimal but practical realizations.
\end{enumerate}

\section{Results and Analysis}\label{sect:simulation}

In this section, the performance of the proposed mmWave channel estimation method employing the beamformers obtained as described above is evaluated numerically.

The simulation scenario is as follows.
Both the transmitter and the receiver are assumed to have the same number of antennas $T \!= \!R \!= \!8$, the number of training beamforming vectors are such that $M_\text{T}M_\text{R} \!=\! 16$, and a sparse channel model as described by equation \eqref{channel_model} with $L\!=\!3$ and $\sigma_{\gamma}^2=1$ was considered, with the \ac{BS}/\ac{UE} antenna subsystems assumed to be uniformly spaced linear antenna arrays, such that 
\begin{equation}
\bv{a}_\text{r}(\phi_l^\text{r}) \!=\! \frac{1}{\sqrt{R}} \big[1, e^{j\frac{2\pi}{\lambda}d\, {\rm{sin}}(\phi_l^\text{r})},\cdots,e^{j\frac{2\pi}{\lambda}(R-1)d\, {\rm{sin}}(\phi_l^\text{r})}\big]^{\text{T}}\!\!,
\label{aresprx}
\vspace{-1ex}
\end{equation}
\begin{equation}
\bv{a}_\text{t}(\phi_l^\text{t}) \!=\! \frac{1}{\sqrt{T}} \big[1, e^{j\frac{2\pi}{\lambda}d\, {\rm{sin}}(\phi_l^\text{t})},\cdots,e^{j\frac{2\pi}{\lambda}(T-1)d\, {\rm{sin}}(\phi_l^\text{t})}\big]^{\text{T}}\!\!,
\label{aresptx}
\end{equation}
with an inter-antenna spacing $d$ of half transmission wavelength $\lambda$, \textit{i.e.} $d \!=\! \lambda/2$, and \ac{AoA}/\ac{AoD} uniformly and randomly distributed in the interval $[0,2\pi]$.

%
% Looking at the Impact of FRAME Design Approach
%

\begin{figure}[H]
\centering
\includegraphics[width=\columnwidth]{\figsdir/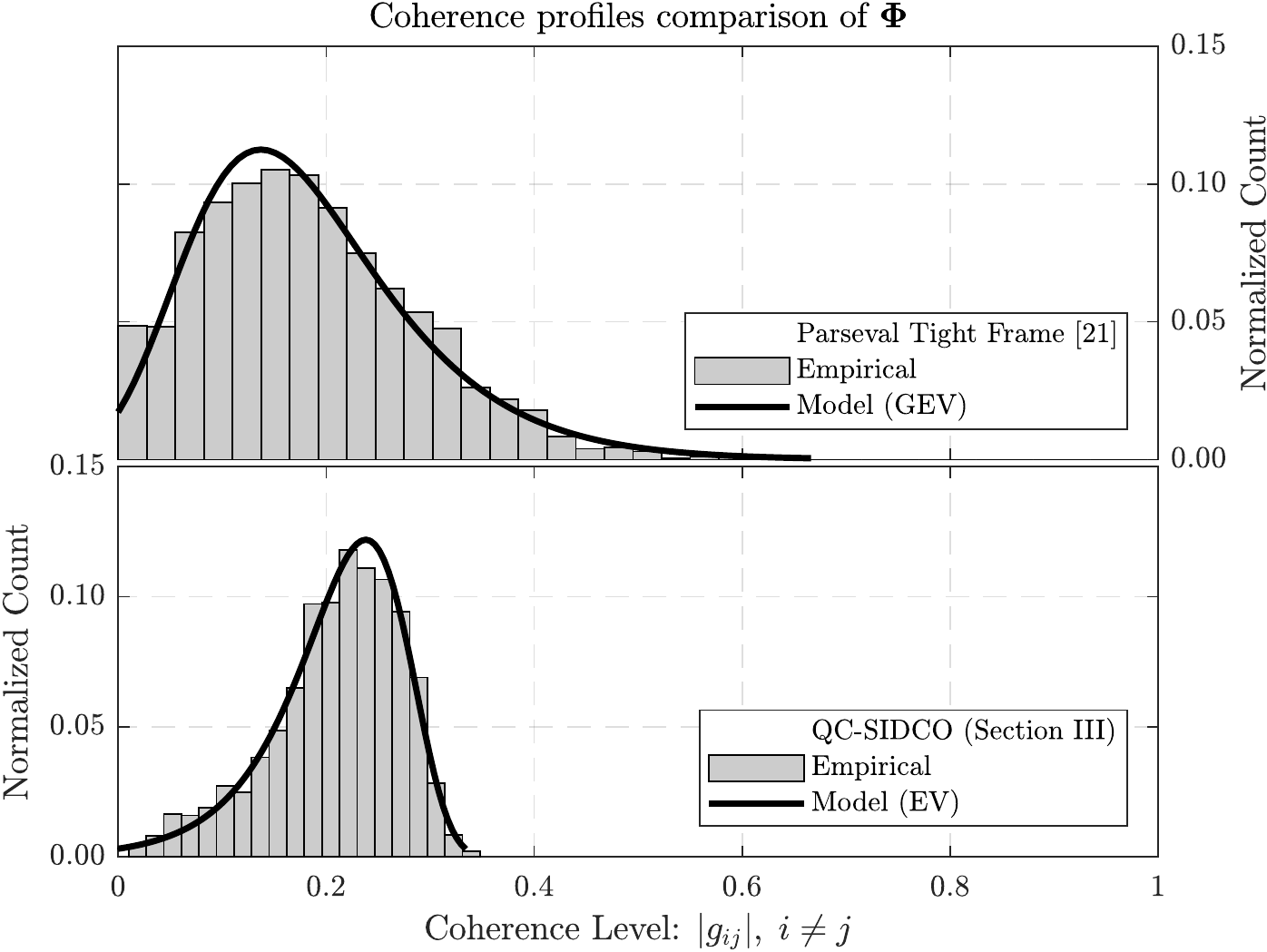}
 \caption[]{Coherence profiles of measurement matrices $\bv{\Phi}$ obtained with the \ac{PTF} construction approach of \cite{chen2013projectiondesignframescs}, and with the low-coherence \ac{QC-SIDCO} frame construction approach of Section \ref{sect:frames_basics}, with $T\! =\! R\! = \! 8$ and $M_\text{T} \!=\! M_\text{R} \!=\! 4$.}\label{fig:pdf_comp}
\vspace{-2ex}
\end{figure}

Let us start our numerical evaluation of the proposed art by studying the impact of the proposed \ac{QC-SIDCO} measurement matrix design.
To this end, we first compare in Fig. \ref{fig:pdf_comp} the \emph{coherence profile} -- defined as the distribution of inner-products $\vert g_{ij}\vert \triangleq \| \boldsymbol{\Phi}_i^{\text{H}}\boldsymbol{\Phi}_j \|_2$ for all distinct column pairs $(i,j)$ -- corresponding to measurement matrices $\bv{\Phi}$ obtained with the \ac{PTF} construction approach of \cite{chen2013projectiondesignframescs}, against that achieved by the low-coherence \acf{QC-SIDCO} frames constructed as described in Section \ref{sect:frames_basics}.
The empirical realizations have been fitted by families of \ac{GEV}, and respectively, \ac{EV} distributions \cite{resnick2013extreme}, \textit{i.e} Fig. \ref{fig:pdf_comp}.

It can be observed that indeed the proposed \ac{QC-SIDCO} approach is superior as it both reduces the frame coherence as defined in equation \eqref{eq:mutual_coh_defi}, yet also preserves the frame tightness, as defined in equation \eqref{eq:frame_defi}.

\vspace{-2ex}

%
% Looking at the Final Estimation Accuracy
%

Next we compare the performance of the \ac{CS}-based channel estimation algorithms employing the ideal measurement matrices highlighted above.
To this end, a grid granularity of $G_\text{T} \!=\! G_\text{R} \!=\! 10$ was considered, and different methods were used to solve the channel estimation problem.
In particular, the classical \ac{OMP} algorithm of \cite{TroppOMP2007} was used to solve equation \eqref{eq_l0combinatorial}, and both the standard \ac{BPDN} algorithm of \cite{donoho1994basispursuit} as well as our previously proposed \ac{BPDN}-$\ell_1$ variation given in \cite[Alg. 1]{malla2016iswcschannel} (with maximum $t = 4$ iterations and tolerance $\epsilon = 0.1$) were used to solve equation \eqref{eq_bpdn}.

The \ac{NMSE} $\mathbb{E} \big[\frac{\|\bv{H} - \hat{\bv{H}}\|_{F}}{\|\bv{H}\|_{F}}\big]$ was used as accuracy measure to compare the performance of the different mmWave channel estimation end schemes.

\vspace{-2ex}
\begin{figure}[H]
\centering
\begin{subfigure}[H]{\columnwidth}
\includegraphics[width=\textwidth]{\figsdir/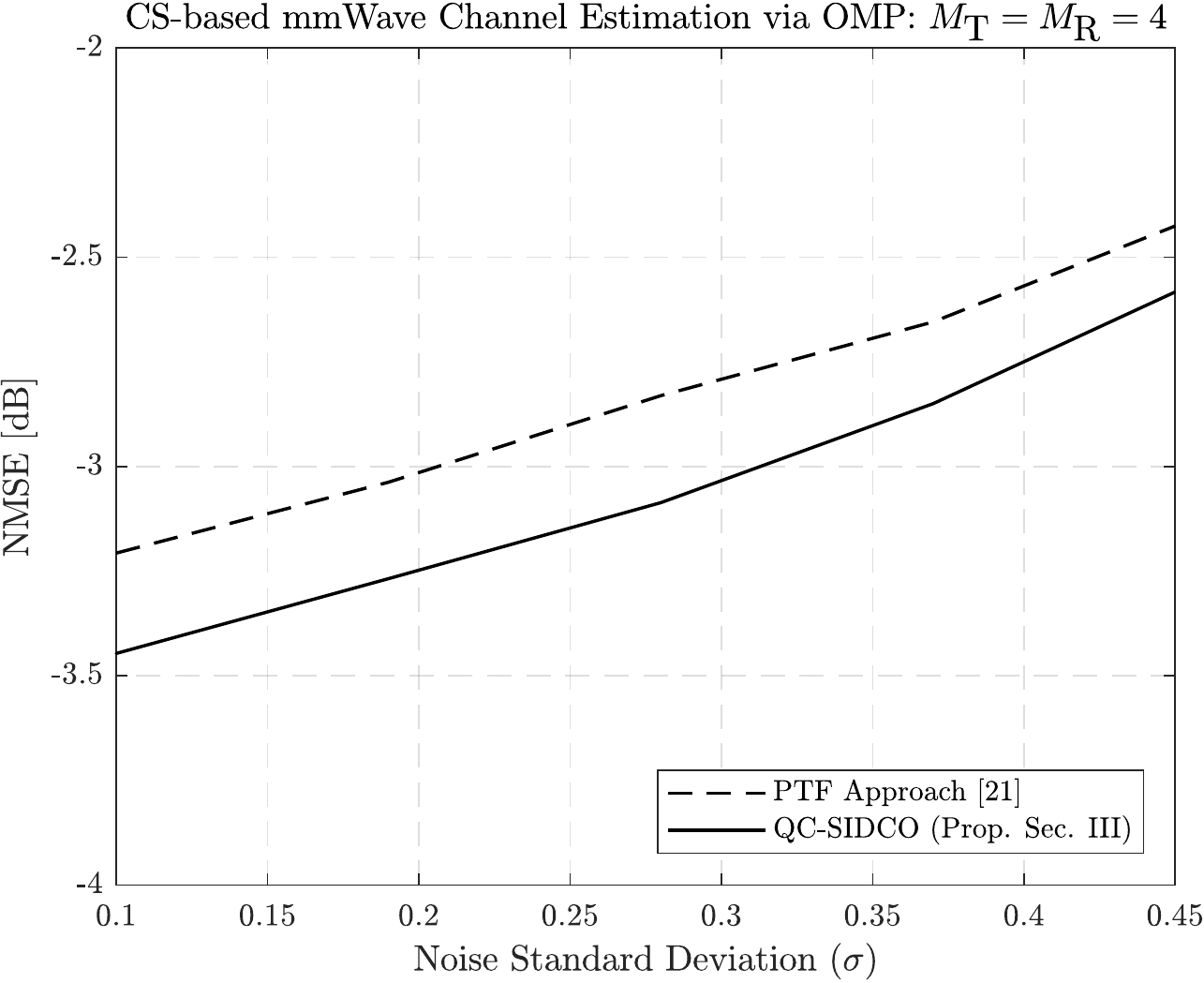}
\caption[]{Channel estimation via equation \eqref{eq_l0combinatorial} solved by \ac{OMP} \cite{TroppOMP2007}.}
\label{fig:comparison_omp}
\end{subfigure}
\vspace{1ex}
\begin{subfigure}[H]{\columnwidth}
\includegraphics[width=\textwidth]{\figsdir/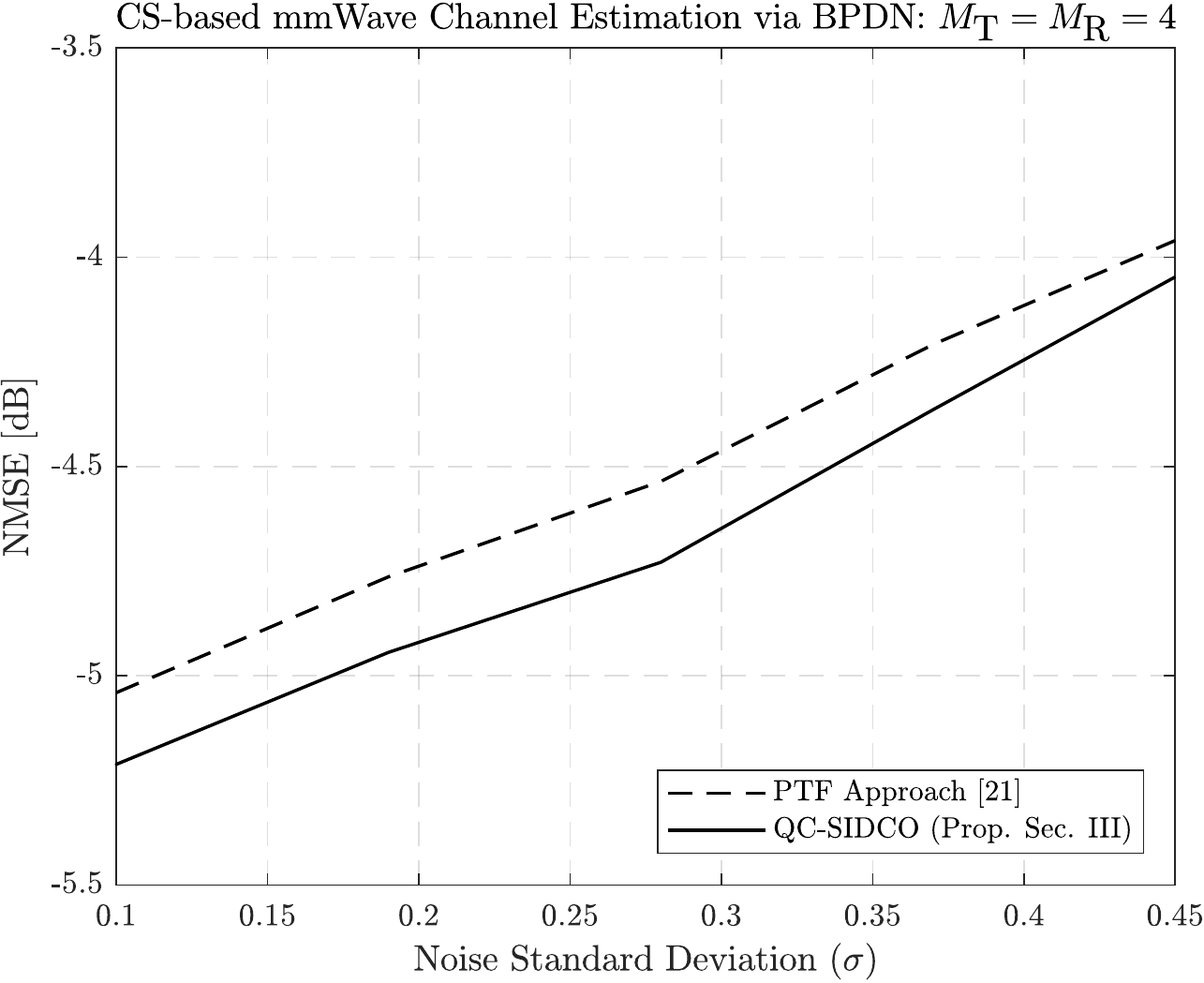}
\caption[]{Channel estimation via equation \eqref{eq_bpdn} solved by \ac{BPDN} \cite{donoho1994basispursuit}.}
\label{fig:comparison_bpdn}
\end{subfigure}
\caption[]{Comparison of \ac{CS}-based mmWave channel estimation algorithms employing measurement matrices obtained with the \ac{PTF} construction approach of \cite{chen2013projectiondesignframescs}, and with the low-coherence \ac{QC-SIDCO} frame method of Section \ref{sect:frames_basics}.}
\label{fig:comparison_omp_bpdn}
\vspace{-2ex}
\end{figure}

% \newpage

In Fig. \ref{fig:comparison_omp_bpdn}, it can be seen that employing the measurement matrix proposed in Section \ref{sect:frames_basics} results in an improved estimation accuracy  compared to the alternative of employing the state-of-the-art \ac{PTF} construction approach of \cite{chen2013projectiondesignframescs}, regardless of whether the estimation problem itself is performed via OMP or BPDN, corresponding to equations \eqref{eq_l0combinatorial} and \eqref{eq_bpdn}, respectively.

\begin{figure}[H]
\centering
\includegraphics[width=\columnwidth]{\figsdir/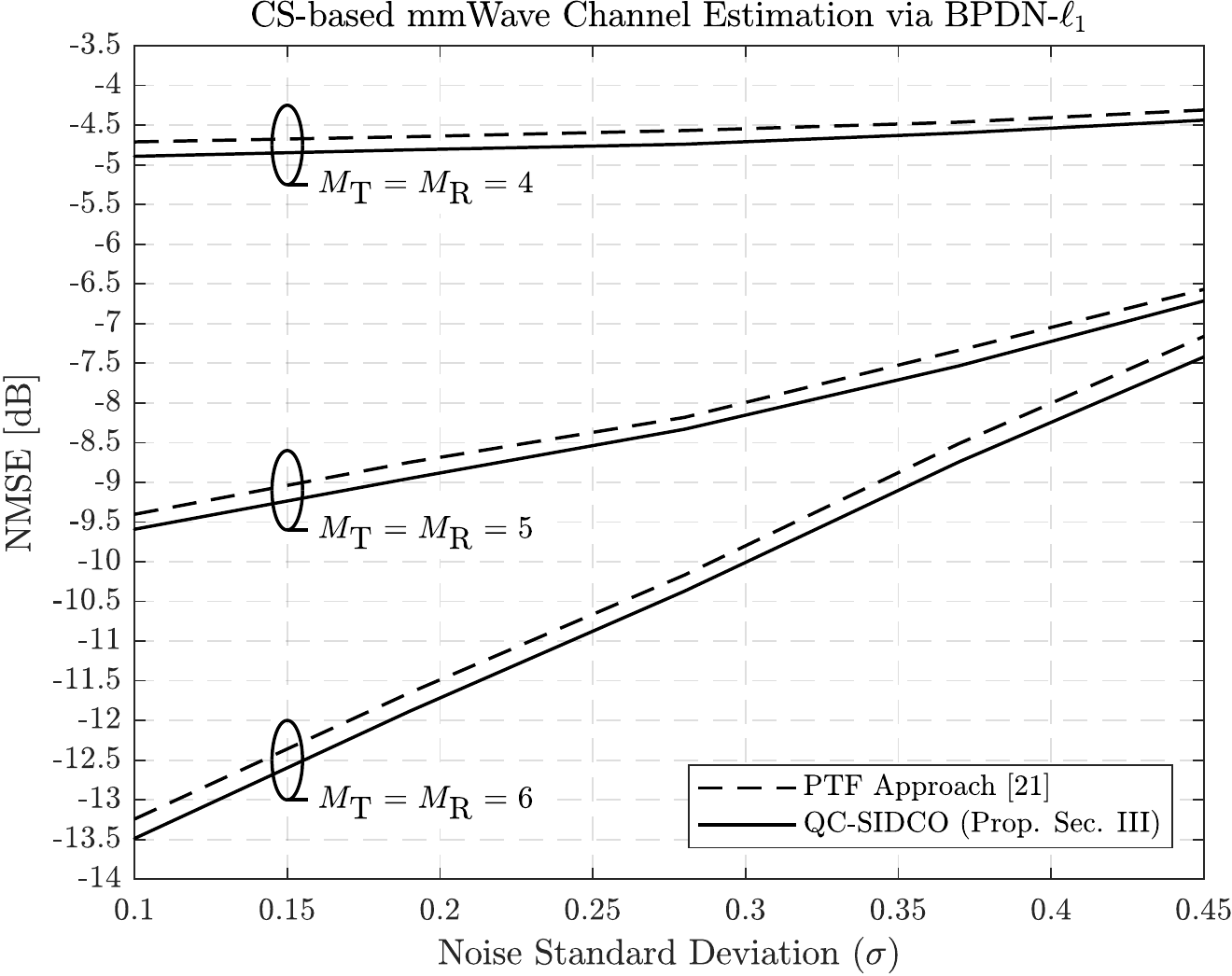}
\caption[]{Sparse recovery performance given \ac{BPDN}-$\ell_1$ algorithm \cite{malla2016iswcschannel} -- a comparison between the \ac{PTF} optimized sensing matrix $\bv{\Omega}$ and the proposed low-coherence \ac{QC-SIDCO} tight frame measurement matrix $\bv{\Phi}$ designs.}\label{fig:csidco_ptf_performance}
\vspace{-2ex}
\end{figure}

Finally, in Fig. \ref{fig:csidco_ptf_performance}, the performance of mmWave channel estimation schemes based on the improved iterative \ac{BPDN} $\ell_1$-reweighed sparse estimator proposed in \cite{malla2016iswcschannel} and employing \ac{PTF} and \ac{QC-SIDCO} is evaluated.

In this last comparison, however, we let $M_\text{T}, M_\text{R}$ assume different values while maintaining $TR$ constant.
Referring to equations \eqref{eq_ChannelSparse2} and \eqref{reconstruction_beamformers_rank1}, it can be seen that this impacts on the aspect ratio of the measurement matrices ($i.e.$, number of rows divided by number of columns), such that the results capture the robustness of the beamforming designs.
Not only is the superiority of the \ac{QC-SIDCO} method of Section \ref{sect:frames_basics} over the \acf{PTF} approach of \cite{chen2013projectiondesignframescs} once again confirmed, but also it is found that the relative gain obtained is robust against the noise power and the frame's aspect ratio. 

\section{Conclusion}
\label{sect:conclusion}
\vspace{-1ex}
We discussed in this paper the problem of sparse recovery of \ac{mmWave}-\ac{MIMO} channels.
In particular, we focused on the design of the training vectors for sparse channel recovery.
We reformulated the problem in the context of Frame Theory and we proposed a novel tight low-coherence generic design associated with the Kronecker product of the transmit and receive beamformers.
We also introduced a recovery mechanism for practical realizations based on the beamforming matrices by means of a vectorized \ac{SVD} decomposition. 
Furthermore, we analyzed the proposed scheme against the state of the art and outlined its design advantages and robust performance gains for the estimation of \ac{mmWave}-\ac{MIMO} channels.

\bibliographystyle{IEEEtran}
% Ignore errors thrown by references section before editing
\vspace{-1ex}
\bibliography{IEEEabrv,\myreferences}

\begin{thebibliography}{10}
\providecommand{\url}[1]{#1}
\csname url@samestyle\endcsname
\providecommand{\newblock}{\relax}
\providecommand{\bibinfo}[2]{#2}
\providecommand{\BIBentrySTDinterwordspacing}{\spaceskip=0pt\relax}
\providecommand{\BIBentryALTinterwordstretchfactor}{4}
\providecommand{\BIBentryALTinterwordspacing}{\spaceskip=\fontdimen2\font plus
\BIBentryALTinterwordstretchfactor\fontdimen3\font minus
  \fontdimen4\font\relax}
\providecommand{\BIBforeignlanguage}[2]{{%
\expandafter\ifx\csname l@#1\endcsname\relax
\typeout{** WARNING: IEEEtran.bst: No hyphenation pattern has been}%
\typeout{** loaded for the language `#1'. Using the pattern for}%
\typeout{** the default language instead.}%
\else
\language=\csname l@#1\endcsname
\fi
#2}}
\providecommand{\BIBdecl}{\relax}
\BIBdecl

\bibitem{RappaportTamir2013}
T.~S. Rappaport, F.~Gutierrez, E.~Ben-Dor, J.~N. Murdock, Y.~Qiao, and J.~I.
  Tamir, ``{Broadband Millimeter-Wave Propagation Measurements and Models Using
  Adaptive-Beam Antennas for Outdoor Urban Cellular Communications},''
  \emph{{IEEE Transactions on Antennas and Propagation}}, vol.~61, no.~4, pp.
  1850--1859, April 2013.

\bibitem{Zorzi2015}
{H. Shokri-Ghadikolaei and C. Fischione and G. Fodor and P. Popovski and M.
  Zorzi}, ``{Millimeter Wave Cellular Networks: A {MAC} Layer Perspective},''
  \emph{{IEEE Trans. on Communications}}, vol.~63, no.~10, pp. 3437--3458, Oct
  2015.

\bibitem{GaoHeath2016}
X.~Gao, L.~Dai, S.~Han, C.~L. I, and R.~W. Heath, ``{Energy-Efficient Hybrid
  Analog and Digital Precoding for MmWave {MIMO} Systems With Large Antenna
  Arrays},'' \emph{{IEEE Journal on Selected Areas in Communications}},
  vol.~34, no.~4, pp. 998--1009, April 2016.

\bibitem{SunRappaport2014}
S.~Sun, T.~S. Rappaport, R.~W. Heath, A.~Nix, and S.~Rangan, ``{{MIMO} for
  Millimeter-wave Wireless Communications: Beamforming, Spatial Multiplexing,
  or Both?}'' \emph{{IEEE Communications Magazine}}, vol.~52, no.~12, pp.
  110--121, December 2014.

\bibitem{ZhouOhashi2014}
L.~Zhou and Y.~Ohashi, ``{Low Complexity Millimeter-wave {LOS-MIMO} Precoding
  Systems for Uniform Circular Arrays},'' in \emph{{IEEE Wireless
  Communications and Networking Conference}}, April 2014, pp. 1293--1297.

\bibitem{LiangDong2014}
L.~Liang, W.~Xu, and X.~Dong, ``{Low-Complexity Hybrid Precoding in Massive
  Multiuser {MIMO} Systems},'' \emph{{IEEE Wireless Communications Letters}},
  vol.~3, no.~6, pp. 653--656, Dec 2014.

\bibitem{Heath_ChEst2014}
A.~Alkhateeb, O.~E. Ayach, G.~Leus, and R.~W. Heath, ``{Channel Estimation and
  Hybrid Precoding for Millimeter Wave Cellular Systems},'' \emph{{IEEE Journal
  of Selected Topics in Signal Processing}}, vol.~8, no.~5, pp. 831--846, Oct
  2014.

\bibitem{Lee2014Globecomm}
J.~Lee, G.~T. Gil, and Y.~H. Lee, ``{Exploiting Spatial Sparsity for Estimating
  Channels of Hybrid {MIMO} Systems in Millimeter Wave Communications},'' in
  \emph{{IEEE Global Communications Conference}}, Dec 2014, pp. 3326--3331.

\bibitem{GhauchEst2015}
H.~Ghauch, M.~Bengtsson, T.~Kim, and M.~Skoglund, ``{Subspace Estimation and
  Decomposition for Hybrid Analog-digital Millimetre-wave {MIMO} Systems},'' in
  \emph{{IEEE International Workshop on Signal Processing Advances in Wireless
  Communications (SPAWC)}}, June 2015, pp. 395--399.

\bibitem{RialITA2015}
R.~M{\'e}ndez-Rial, C.~Rusu, A.~Alkhateeb, N.~Gonz{\'a}lez-Prelcic, and R.~W.
  Heath, ``{Channel Estimation and Hybrid Combining for mmWave: Phase Shifters
  or Switches?}'' in \emph{{Information Theory and Applications Workshop (ITA),
  2015}}, Feb 2015, pp. 90--97.

\bibitem{Montagner2015}
S.~Montagner, N.~Benvenuto, and P.~Baracca, ``{Channel Estimation Using a 2{D}
  {DFT} for Millimeter-Wave Systems},'' in \emph{{2015 IEEE 81st Vehicular
  Technology Conference (VTC Spring)}}, May 2015, pp. 1--5.

\bibitem{NiDong2016}
W.~Ni and X.~Dong, ``{Hybrid Block Diagonalization for Massive Multiuser MIMO
  Systems},'' \emph{{IEEE Transactions on Communications}}, vol.~64, no.~1, pp.
  201--211, Jan 2016.

\bibitem{MyListOfPapers:HeathJSASP2016}
R.~W. Heath, N.~Gonz{\'a}lez-Prelcic, S.~Rangan, W.~Roh, and A.~M. Sayeed,
  ``{An Overview of Signal Processing Techniques for Millimeter Wave MIMO
  Systems},'' \emph{{{IEEE} Journal on Selected Topics on Signal Processing}},
  vol.~10, no.~3, pp. 436 -- 453, 2016.

\bibitem{DonohoCS2006}
D.~L. Donoho, ``{Compressed Sensing},'' \emph{{IEEE Transactions on Information
  Theory}}, vol.~52, no.~4, pp. 1289--1306, April 2006.

\bibitem{TroppOMP2007}
J.~A. Tropp and A.~C. Gilbert, ``{Signal Recovery From Random Measurements Via
  Orthogonal Matching Pursuit},'' \emph{{IEEE Transactions on Information
  Theory}}, vol.~53, no.~12, pp. 4655--4666, Dec 2007.

\bibitem{malla2016iswcschannel}
S.~Malla and G.~Abreu, ``{Channel Estimation in Millimeter Wave MIMO Systems:
  Sparsity Enhancement via Reweighting},'' in \emph{{2016 International
  Symposium on Wireless Communication Systems (ISWCS)}}, Sept 2016, pp.
  230--234.

\bibitem{donoho1994basispursuit}
S.~Chen and D.~Donoho, ``{Basis Pursuit},'' in \emph{{Proceedings of 1994 28th
  Asilomar Conference on Signals, Systems and Computers}}, vol.~1, Oct 1994,
  pp. 41--44 vol.1.

\bibitem{CandesBoydL12007}
E.~J. Cand{\`e}s, M.~B. Wakin, and S.~P. Boyd, ``{Enhancing Sparsity by
  Reweighted $\ell_1$ Minimization},'' \emph{{Journal of Fourier Analysis and
  Applications}}, vol.~14, no.~5, pp. 877--905, Dec 2008.

\bibitem{casazza2012finite}
P.~G. Casazza and G.~Kutyniok, \emph{{Finite Frames: Theory and Applications}},
  1st~ed.\hskip 1em plus 0.5em minus 0.4em\relax {Birkh\"{a}user Basel}, 2012.

\bibitem{rusu2016designing}
C.~Rusu and N.~Gonz{\'a}lez-Prelcic, ``{Designing Incoherent Frames Through
  Convex Techniques for Optimized Compressed Sensing},'' \emph{{IEEE
  Transactions on Signal Processing}}, vol.~64, no.~9, pp. 2334--2344, May
  2016.

\bibitem{chen2013projectiondesignframescs}
{W. Chen, M. R. D. Rodrigues and I. J. Wassell}, ``{Projection Design for
  Statistical Compressive Sensing: A Tight Frame Based Approach},'' \emph{{IEEE
  Transactions on Signal Processing}}, vol.~61, no.~8, pp. 2016--2029, April
  2013.

\bibitem{kovacevic2008introduction}
J.~Kova\u{c}evic and A.~Chebira, ``{An Introduction to Frames},''
  \emph{{Foundations and Trends in Signal Processing}}, vol.~2, no.~1, pp.
  1--94, 2008.

\bibitem{strohmer2003grassmanianapps}
{T. Strohmer and R. W. Heath Jr.}, ``{Grassmannian Frames with Applications to
  Coding and Communication},'' \emph{{Applied Computational Harmonic
  Analysis}}, vol.~14, pp. 257--275, 2003.

\bibitem{donoho2006compressed}
D.~L. Donoho, ``{Compressed Sensing},'' \emph{{IEEE Transactions on Information
  Theory}}, vol.~52, no.~4, pp. 1289--1306, 2006.

\bibitem{tropp2006justrelax}
J.~A. Tropp, ``{Just Relax: Convex Programming Methods for Identifying Sparse
  Signals in Noise},'' \emph{{IEEE Transactions on Information Theory}},
  vol.~52, pp. 1030--1051, 2006.

\bibitem{rusu2018algorithms}
C.~Rusu, N.~Gonzalez-Prelcic, and R.~W. Heath~Jr, ``{Algorithms for the
  Construction of Incoherent Frames Under Various Design Constraints},''
  \emph{arXiv preprint arXiv:1801.09678}, 2018.

\bibitem{tropp2005designing}
J.~A. Tropp, I.~S. Dhillon, R.~W. Heath, and T.~Strohmer, ``{Designing
  Structured Tight Frames via an Alternating Projection Method},'' \emph{{IEEE
  Transactions on Information Theory}}, vol.~51, no.~1, pp. 188--209, 2005.

\bibitem{resnick2013extreme}
S.~I. Resnick, \emph{{Extreme Values, Regular Variation and Point
  Processes}}.\hskip 1em plus 0.5em minus 0.4em\relax Springer, 2013.

\end{thebibliography}
\vspace{-1ex}

\end{document}